\documentclass[journal,comsoc,onecolumn,draftclsnofoot]{IEEEtran}

\usepackage{amsmath}
\usepackage[cmintegrals]{newtxmath}

\usepackage[pdftex]{graphicx}
 
\usepackage{amssymb}
\usepackage{subcaption}
\usepackage{epstopdf}
\usepackage{color}
\usepackage{cite}
\usepackage{amsmath}
\DeclareGraphicsExtensions{.eps}

\usepackage{multirow}

\begin{document}

\title{Performance Enhancement of Downlink NOMA by Combination with GSSK}
\author{Jin Woo Kim, and 
        Soo Young Shin, \textit{Senior Member, IEEE,}\\
		Victor C.M.Leung \textit{Fellow, IEEE}

\thanks{Manuscript received XXX, XX, 2016; revised XXX, XX, 2017.}%
\thanks{This research was supported by Basic Science Research Program through the National Research Foundation of Korea (NRF) funded by the Ministry of Education (2015R1D1A1A01061075).}%
\thanks{Jin Woo Kim and Soo Young Shin are with the WENS Lab., Dept. of IT  Convergence Engineering, Kumoh National Institute of Technology, 39177, Gumi, Republic of Korea. (email:rerua@kumoh.ac.kr, wdragon@kumoh.ac.kr).}%
\thanks{V. Leung is with the Department of Electrical and Computer Engineering, University of British Columbia, Canada (e-mail: vleung@ece.ubc.ca).}}%

\maketitle

\begin{abstract}
In non-orthogonal multiple access (NOMA), cell-edge users experience significantly low spectral density because only some part of the total transmit power is allocated. This leads to low spectral efficiency for the paired users in NOMA. To overcome this problem, we propose an integration of NOMA and generalized space shift keying (GSSK), called NOMA-GSSK, to improve the spectral efficiency by exploiting the spatial domain. Spectral and energy efficiency, bit error rate (BER), and computational complexity of the proposed system were analyzed and compared to those of multiple-input multiple-output NOMA (MIMO-NOMA). It is shown that NOMA-GSSK outperforms MIMO-NOMA.
\end{abstract}


\IEEEpeerreviewmaketitle

\section{Introduction}

\IEEEPARstart{I}{n} recent years, the amount of network traffic has increased significantly because of the large number of users connecting to the network. Moreover, the boom in the Internet of Things is expected to increase network traffic dramatically. To address this soaring traffic demand, next-generation wireless technologies such as 5G are required to provide advantages, such as better spectral efficiency, massive connectivity, and faster response time \cite{1}.
\par
Non-orthogonal multiple access (NOMA) is a promising candidate for 5G to achieve better capacity gains because of its high spectral efficiency \cite{2,3,4}. In \cite{2}, the author classified NOMA as code domain and power domain. In this letter, we aimed for power domain NOMA (hereinafter referred to as NOMA). In NOMA, multiple users are served in each orthogonal resource block, e.g., a time slot, frequency channel, or spreading code, by exploiting the power domain. The signals of multiplexed users are allocated different power levels by the base station (BS), superimposed with each other, and transmitted. To recover their signals, cell center users perform successive interference cancellation (SIC) \cite{5}. However, cell-edge users do not perform SIC, and experience a decrease in spectral efficiency due to degraded signals. Recently, to enhance the spectral efficiency, NOMA with spatial modulation (SM) was investigated in \cite{wang2017achievable}, \cite{wang2017spectral}. In those studies, the authors aimed to analyze the spectral efficiency of SM-NOMA from the point of view of mutual information.
\par
To solve the problem of cell-edge users, NOMA-SSK has been suggested, in which the cell-edge user is multiplexed in the spatial domain to improve the spectral efficiency of the system by using NOMA and space shift keying (SSK) \cite{6}. SSK is a multiple-input multiple-output (MIMO) technique, which transmits information using an antenna index, contrary of the conventional modulation schemes \cite{7}. Moreover, the application of SSK can efficiently reduce transmitter overhead and receiver complexity by using the antenna index instead of any modulation scheme. However, because of the characteristics of SSK, the number of transmit antennas must be a power of two.
\par
In this letter, to further improve the spectral efficiency of cell-edge users and to overcome the limitation on the number of transmit antennas of NOMA-SSK, we propose a novel transmission scheme by combining NOMA and generalized space shift keying (GSSK), called NOMA-GSSK. GSSK is a generalized form of SSK and uses multiple transmit antennas, unlike SSK \cite{8}. The proposed scheme achieves higher spectral and energy efficiency, and lower bit error rate (BER), compared to MIMO-NOMA and NOMA-SSK, because the users are multiplexed in both the power and spatial domains, by using a set of transmit antennas, whereas in MIMO-NOMA, all antennas are used to transmit the NOMA signal. In addition, because cell-edge users are multiplexed in the spatial domain, the complexity of the system is also decreased, because the SIC steps are reduced.
\par

\section{System Model}
\begin{figure}[b]
\centering
\includegraphics[width=0.27\textwidth]{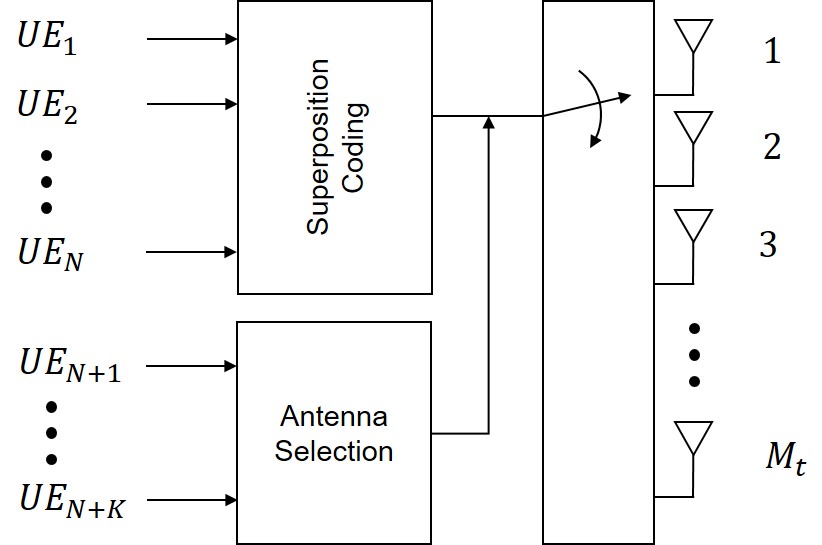}
\caption{NOMA-GSSK Transmitter model.}
\label{fig:Transmit_Model}
\end{figure}

Assume $N+K$ users to be uniformly distributed in a cell,
$N$ users are multiplexed using NOMA, and $K$ users exploit the spatial domain. The channel gains of users are in the order $|h_1| \geq \cdots \geq |h_{N}| \geq |h_{N+1}| \geq \cdots \geq |h_{N+K}|$. The users with low channel gain are regarded as cell-edge users, and are multiplexed by GSSK. Fig.\ref{fig:Transmit_Model} shows the downlink transmitter model for $K$ users. Fractional transmit power allocation (FTPA) is used to allocate ower to the $N$ NOMA users. The transmitted signal is
\begin{equation}
X=\sum_{i=1}^{N}\sqrt{\alpha_i P}x_i,
\end{equation}
where $\alpha_i$ is the $i_{th}$ user's power allocation factor, such that $\alpha_1<\alpha_2<\cdots<\alpha_i<\cdots<\alpha_{N}$ and $\sum_{i=1}^{N}\alpha_i=1$, $P$ is the total transmit power, and $x_i$ is the symbol of the $i$-th user.
\par
As shown in Fig.\ref{fig:Transmit_Model}, $N$ users are transmitted using specific antennas from the entire set of transmit antennas, $M_t$. The data symbols of cell-edge users, $UE_{K}$, are transmitted by using the selected specific antenna set on the basis of the antenna index information. Symbols for the cell-edge users are transmitted by the antenna allocation on the basis of the GSSK mapping rule. For the example of $M_{a}=2$ active transmit antennas, transmitted signal $X_i$ is expressed as
\begin{equation}
X_i \triangleq [\dfrac{X}{\sqrt{M_a}}\;\;\;\cdots\;\;\;0\;\;\;\dfrac{X}{\sqrt{M_a}}\;\;\;\cdots\;\;\;0]^T,
\end{equation}
where $M_a$ is number of active transmit antennas.
\par
The received signal of the $i$-th user can be expressed as
\begin{equation}
y_i=h_{i,j}X_i+n_{i,j},
\end{equation}
where $h_{i,j}$ is the channel gain of the $i$-th user using the $j$-th antenna set, and $n_{i,j}$ is additive white Gaussian noise (AWGN).
\begin{figure}[t]
    \centering
    \begin{subfigure}[b]{0.27\textwidth}
        \centering
        \includegraphics[width=1\textwidth]{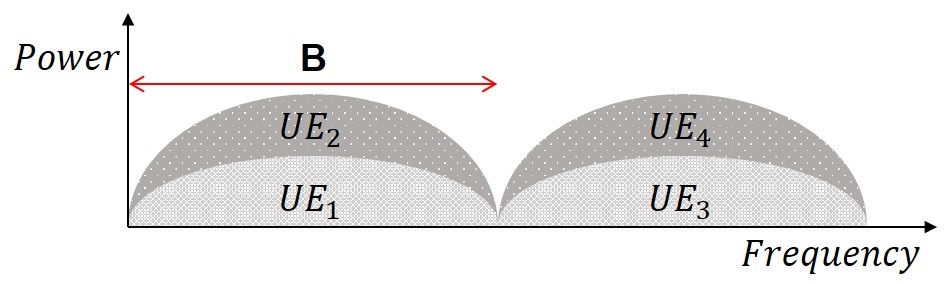}
        \caption{NOMA}
    \end{subfigure}%
    ~

    \begin{subfigure}[b]{0.23\textwidth}
        \centering
        \includegraphics[width=1\textwidth]{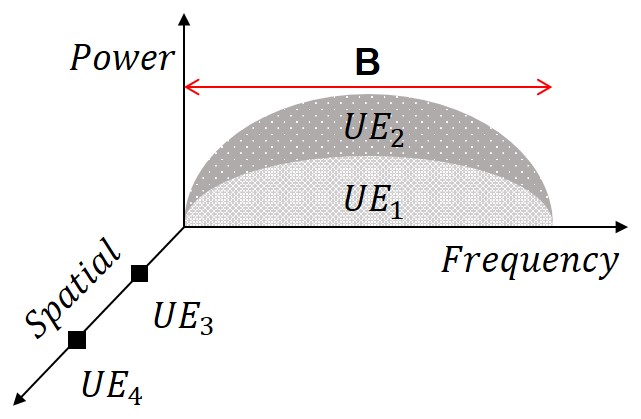}
        \caption{NOMA-GSSK}
    \end{subfigure}
    \caption{Frequency Distribution of NOMA and NOMA-GSSK in 4 users case.}
    \label{fig:Frequency_Band_NOMA_GSSK}
\end{figure}

\par
In the GSSK case, each active transmit antenna sends only a constant signal $1/{\sqrt{M_a}}$, because it transmits only antenna index information based on the set of transmit antennas. However, NOMA-GSSK transmits $X/\sqrt{M_a}$ symbols , similar to generalized spatial modulation (GSM) \cite{9}. By transmitting the superposed signal $X_i$ and antenna index information together, spectral efficiency is improved. 
\par
$N$ NOMA users detect transmitted signals in the same way as NOMA. However, because of the characteristics of NOMA-GSSK, the detection method of the received data is different from that of NOMA. NOMA users are detected by SIC, and $K$ users multiplexed in the spatial domain are detected by a maximum likelihood (ML) detector. Moreover, because only $N$ users are multiplexed in the power domain, the complexity of the system can be decreased by reducing the use of SIC, which requires high complexity.
\par
For example, in the case of $4$ users ($N+K=4$,), Fig. \ref{fig:Frequency_Band_NOMA_GSSK} shows the comparison of frequency distribution between NOMA and NOMA-GSSK. In the example, we assume that N=2 users can be multiplexed by NOMA.
Unlike NOMA, in NOMA-GSSK, $UE_3$ and $UE_4$ are multiplexed in the spatial domain, and the remaining users in the power domain, i.e., NOMA. Assuming that one channel bandwidth is $15$~kHz (LTE's sub-band channel bandwidth), NOMA requires $30$~kHz for $4$ users, considering two users per channel. NOMA-GSSK can support all $4$ users with only $15$~kHz ($1$ sub-channel). This shows that NOMA-GSSK has naturally better spectral efficiency than NOMA. There are $N_H$ possible index sets $j$ representing the active antennas, i.e., $b_H$ = $log2(N_H)$ bits can be conveyed by the particular choice of index set j. If K$>$1 cell-edge users are supposed to be supported, they have to share these $b_H$ bits, i.e., each user will receive $b_H/K$ bits-per-channel-use (bpcu).
\subsection{Cell-edge user detector}
In the proposed NOMA-GSSK, GSSK is used for the symbol transmission of cell-edge users, and symbol detection is performed by determining which set of transmit antennas is actively transmitting. Cell-edge users receive NOMA symbols, and evaluate the transmit antenna index used at the BS.
Received signals are demodulated by using an ML detector. For each $i_{th}$ cell-edge user, ML detection can be expressed as
\begin{equation}
\hat{l}=\underset{\textbf{j}}{\arg\min}\Vert y_i-\sqrt{\rho'}h_{j,eff} \Vert ^2,
\end{equation}
where $\hat{l}$ is the detected transmit antenna set, $\rho'$ is signal-to-noise ratio (SNR), $h_{j,eff}$ is the effective channel gain of the $j_{th}$ antenna set ($h_{j,eff}=h_{j(1)}+h_{j(2)}+\cdots+h_{j(M_a)}$, \textbf{j}($\cdot$) = $j \in \{1,2,...,M_t\}$). It is to be noted that, as the cell-edge users information is modulated using antenna set, they are only concerned about the transmit antenna set detection.
Error performance of the ML detector can be derived as

\setlength{\arraycolsep}{0.0em}
\begin{eqnarray}
P_e \leq \dfrac{1}{N_H log_2 (N_H)} \sum_{j}^{N_H} \sum_{k,k \neq j}^{N_H} N_{b_{j,k}} Q \left( A \right),\\
A = \sqrt{\dfrac{\overline{\gamma}}{M_a} \vert \sum_{l=1}^{M_a}[h_{j(l)}-h_{k(l)}] \vert^2 },
\end{eqnarray}
where $N_{b_{j,k}}$ is the number of error bits between the $j$-th and $k$-th constellation points that follow the GSSK mapping rule, $N_H$ the possible constellation with size of a power of 2 ($j=1,2,...,N_H$), $Q(\cdot)=1/\sqrt{2\pi}\int_{0}^{\infty}exp(-u^2/2)du$, $\overline{\gamma}$ is the SNR (average SNR), and $h_{x(l)}$ is the $x$-th constellation point \cite{8}. 
\par
The sum-rate of all cell-edge users ($UE_{N+1}$,$\cdots$,$UE_{N+K}$) is expressed as
\begin{equation}
R_{K}=(1-P_e) \lfloor \log_2(_{M_t}C_{M_a}) \rfloor,
\end{equation}
where $_{M_t}C_{M_a}$ is the binomial coefficient of ($M_t$, $M_a$).
\section{Performance Analysis}
\subsection{Capacity Analysis}
For a total of $N+K$ users in a MIMO-NOMA system\cite{4}, the capacity is given by
\begin{equation}
R_{MIMO-NOMA}=\log_2(\rho \log_2(\log_2{(N+K)})).
\end{equation}
\par
The capacity of NOMA-SSK and NOMA-GSSK can be calculated as the sum of the capacity of $N$ NOMA users and the capacity of cell-edge users using the spatial domain. NOMA-SSK has an average capacity given as
\begin{equation}
R_{NOMA-{SSK}}=\log_2(\rho \log_2(\log_2 (N)))+(1-P_e)\lfloor \log_2(M_t) \rfloor.
\end{equation}
\par
The capacity of NOMA-GSSK is given by
\begin{equation}
R_{NOMA-{GSSK}} = \log_2(\rho \log_2(\log_2 (N)))+R_K.
\end{equation}
\par

Fig. \ref{fig:SE_diff_1} shows the capacity comparison with respect to the number of transmit antennas. When the number of transmit antennas is less than $4$, the capacity of NOMA-SSK is equal to that of NOMA-GSSK. However, when the number of transmit antennas is more than $4$, NOMA-GSSK has a higher capacity, because GSSK can have a plurality of active transmit antenna sets rather than one active transmit antenna.
\subsection{Energy Efficiency Analysis}
Generally, the energy efficiency is expressed as
\begin{equation}
\eta = \dfrac{R}{P_T},
\end{equation}
where $R$ denotes the capacity, and $P_T$ is the total transmit power.
\par
The cell-edge user is multiplexed in the spatial domain and the information is transmitted using the antenna index set. Because the cell-edge user of NOMA-GSSK does not use power allocation, NOMA-GSSK has superior energy efficiency compared to MIMO-NOMA.
\par
In MIMO-NOMA, where the total power is allocated to the entire number of users $N+K$, the total power of MIMO-NOMA can be expressed as $P_{T(MIMO-NOMA)} = \sum_{i=1}^{N+K}\alpha_i P$. NOMA-GSSK assigns the total power to users other than the cell-edge users like NOMA-SSK does. Therefore, $P_{T(NOMA-SSK)} = P_{T(NOMA-GSSK)} = \sum_{i=1}^{N}\alpha_i P$ and the energy efficiency of MIMO-NOMA, NOMA-SSK, and NOMA-GSSK are given by
\begin{equation}
\eta_{MIMO-NOMA} = \dfrac{R_{MIMO-NOMA}}{\sum_{i=1}^{N+K}\alpha_i P},
\end{equation}
\begin{equation}
\eta_{NOMA-SSK} = \dfrac{R_{NOMA-SSK}}{\sum_{i=1}^{N}\alpha_i P},
\end{equation}
\begin{equation}
\eta_{NOMA-GSSK} = \dfrac{R_{NOMA-GSSK}}{\sum_{i=1}^{N}\alpha_i P}.
\end{equation}
\par
Eqs. (12)-(14) clearly show that NOMA-GSSK has improved energy efficiency compared to conventional schemes.
\subsection{Complexity Analysis}
The complexity of SIC can be divided into two parts: decoding and subtraction. In this system, because an ML detector is used, the complexity of MIMO-NOMA in UE$_j$ can be obtained as
\begin{equation}
O_{MIMO-NOMA} = (4M_{r}M_{t}M + 2M_{r}M^{M_{t}})(N+K-j+1),
\end{equation}
where $M_r$ is the number of receive antennas, $M$ is the modulation order, and $j$ is the ordering for UE from the nearest UE $(1 \leq j \leq N+K)$. In (15), $4M_{r}M_{t}M + 2M_{r}M^{M_{t}}$ is the decoding part based on the ML detector \cite{10}, $(N+K-j+1)$ is the subtraction part, and the unit of complexity is the number of add-compare operations.
\par 
Indeed, the subtraction step of UE$_j$ is $N+K-1$, because the last user of NOMA does not perform SIC. However, NOMA users should decode their own signals after subtraction.
\par
From (15), the complexity for all users can be obtained as
\begin{equation}
O_{MIMO-NOMA,total} = \dfrac{(N+K)(2+(N+K-1))}{2}(4M_{r}M_{t}M + 2M_{r}M^{M_{t}}).
\end{equation}
\par
The complexity of NOMA-SSK and NOMA-GSSK can be calculated using the same approach. However, SSK and GSSK decoding complexity is different from that of the MIMO-ML detector. For this reason, we applied the decoding complexity equation from \cite{7}:
\begin{equation}
O_{NOMA-SSK,total} = \dfrac{N(2+(N-1))}{2}(4M_{r}M_{t}M + 2M_{r}M^{M_{t}})N+(KN_{r}M).
\end{equation}
\par
In the NOMA-GSSK case, the number of transmit antennas is less than in the MIMO-NOMA and NOMA-SSK cases for the same number of accommodated users, ($N+K$). Therefore, it can achieve lower complexity than the other schemes.
\begin{equation}
O_{NOMA-GSSK,total} = \dfrac{N(2+(N-1))}{2}(4M_{r}M_{t}M + 2M_{r}M^{M_{t}})N + (KM_{a}M_{r}log_2(_{M_{t}}C_{M_{a}})).
\end{equation}
\par
In Table \ref{complex_table}, with some numerical examples, we show that the complexity of NOMA-GSSK is lower than that of MIMO-NOMA and NOMA-SSK. The low complexity of NOMA-GSSK is one of the advantages that makes it easier to implement than other schemes (e.g., MIMO-NOMA, NOMA-SSK).
\begin{table*}[]
\centering
\caption{Comparison of the complexity of MIMO-NOMA, NOMA-SSK and NOMA-GSSK.}
\label{complex_table}
\begin{tabular}{|c|c|c|c|c|c|c|c|c|c|}
\hline
\multirow{2}{*}{N+K} & \multirow{2}{*}{K} & \multirow{2}{*}{$M_r$} & \multirow{2}{*}{$M_t$ (MIMO-NOMA, NOMA-SSK)} & \multirow{2}{*}{$M_t$ (GSSK)} & \multirow{2}{*}{$M_a$} & \multirow{2}{*}{M} & \multicolumn{3}{c|}{Complexity}              \\ \cline{8-10} 
                   &                    &                     &                                      &                            &                     &                    & MIMO-NOMA            & NOMA-SSK       & NOMA-GSSK \\ \hline
5                  & 2                  & 4                   & 4                                    & 4                          & 1                   & 2                  & 3,840            & 1,024            & 1,024       \\ \hline
5                  & 2                  & 4                   & 8                                    & 5                          & 2                   & 3                  & 793,080         & 317,256        & 5,072     \\ \hline
\end{tabular}
\end{table*}

\begin{figure*}[t!]
        \centering
        \begin{subfigure}[t]{0.22\textwidth}
			\centering
			\includegraphics[height=100pt,width=\textwidth]{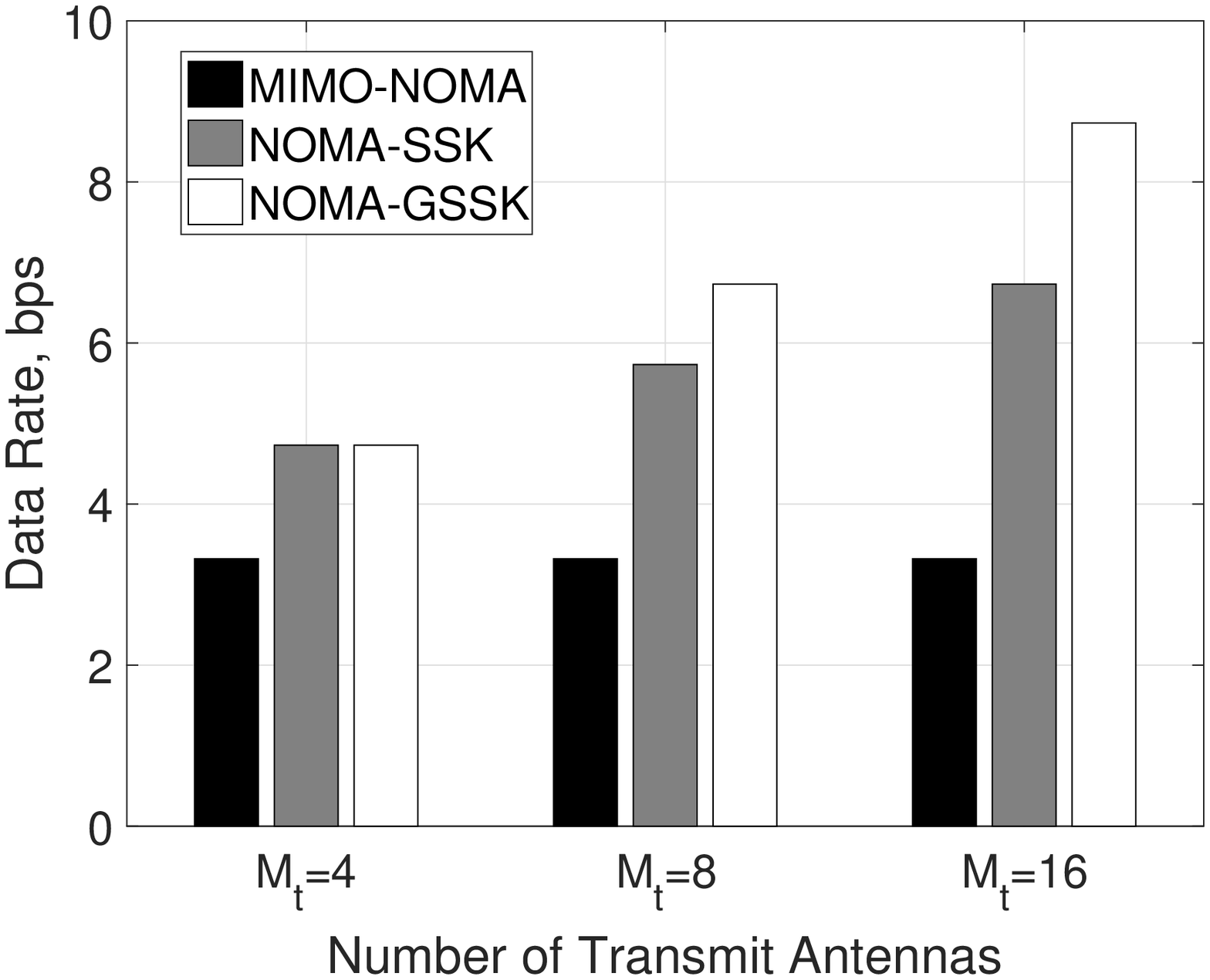}
        	\caption{Data Rate (NOMA-GSSK, $M_a$ = 2).}
        	\label{fig:SE_diff_1}
        \end{subfigure}
        \begin{subfigure}[t]{0.22\textwidth}
            \centering
            \includegraphics[height=100pt,width=\textwidth]{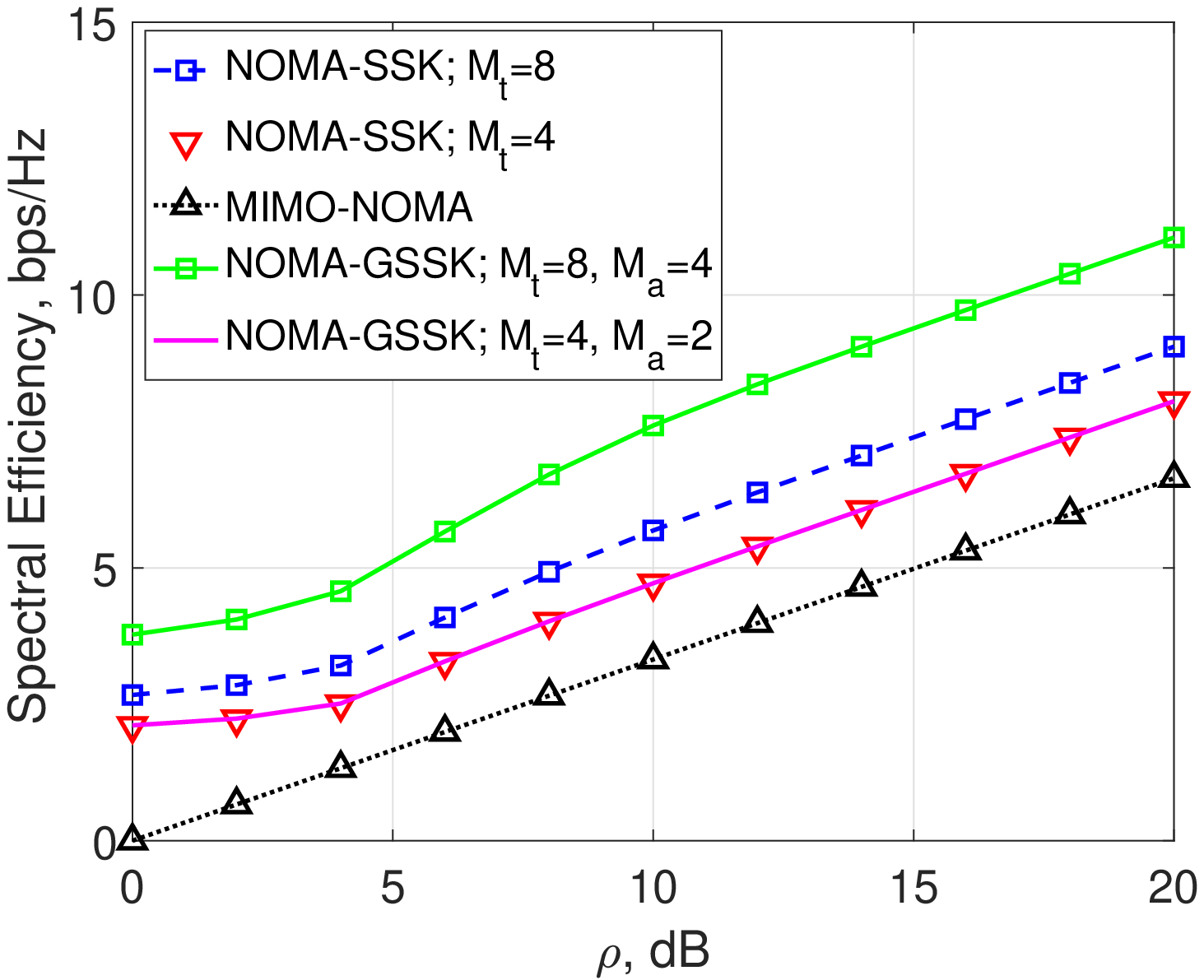}
            \caption{Spectral Efficiency}\label{fig:SE}
        \end{subfigure}
        \begin{subfigure}[t]{0.22\textwidth}
            \centering
            \includegraphics[height=100pt,width=\textwidth]{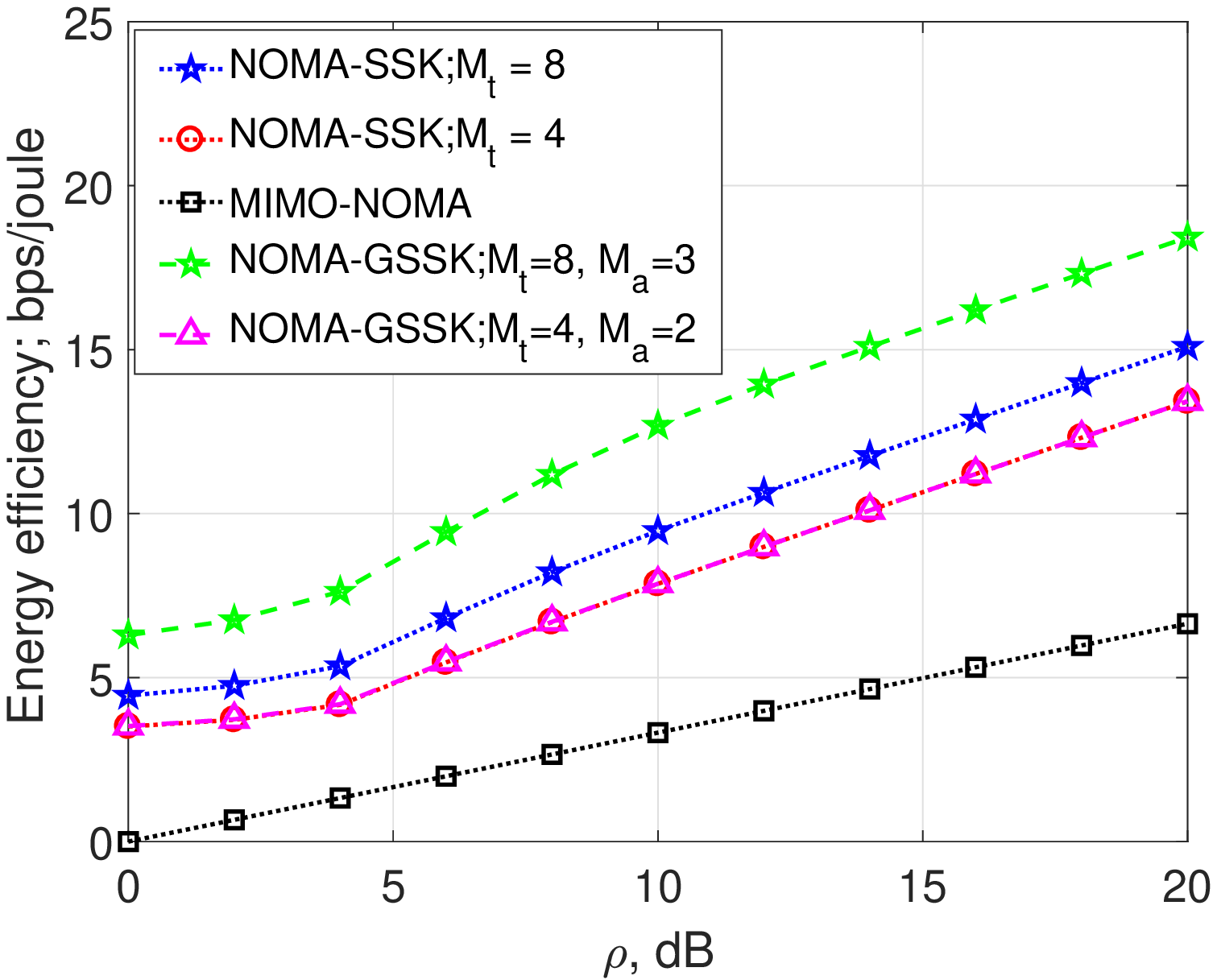}
            \caption{Energy Efficiency}\label{fig:EE}
        \end{subfigure}
        \begin{subfigure}[t]{0.22\textwidth}
            \centering
            \includegraphics[height=100pt,width=\textwidth]{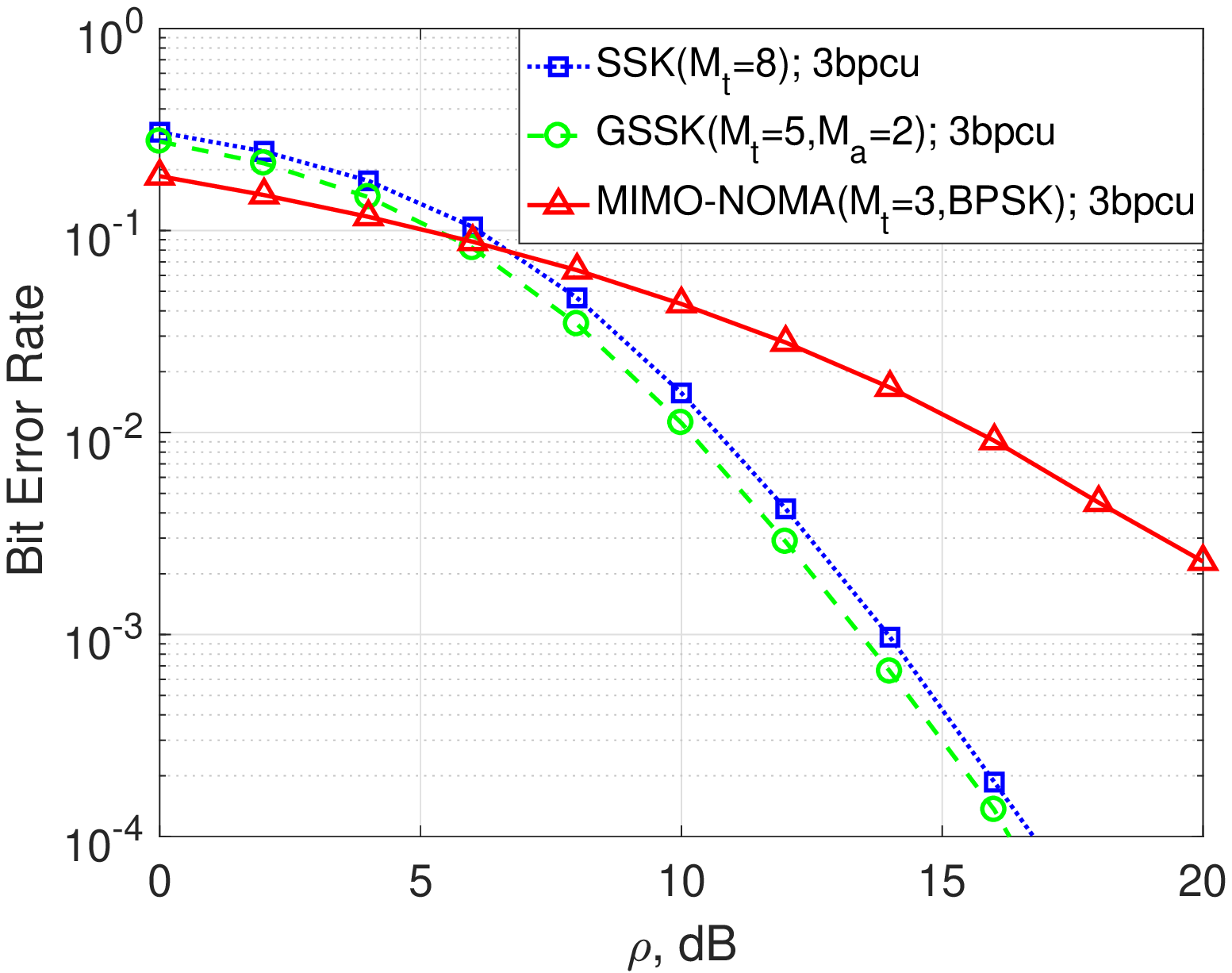}
            \caption{BER of cell-edge user}\label{fig:BER}
        \end{subfigure}
        \caption{Performance comparison of MIMO-NOMA, NOMA-SSK, and NOMA-GSSK.}\label{fig:total}
\end{figure*}

\section{Numerical Results} 
This section discusses the evaluation of the performance of the proposed scheme by simulation. We assumed that the receiver perfectly knows the channel state information (CSI) of the flat Rayleigh fading channel with AWGN. We let $M_r=4$, $N=2$, $K=1$ in all simulations (MIMO-NOMA, $N=3$, $K=0$). In Fig.~\ref{fig:SE_diff_1} and \ref{fig:SE}, $M_t=2$ for MIMO-NOMA. User channel gains are in the range $0 \leq |h_i| \leq 1$. For cell-center users, $|h_i| \geq 0.6$, whereas for cell-edge users, $|h_i| \leq 0.4.$
\par
Fig.~\ref{fig:SE} compares the spectral efficiency of MIMO-NOMA, NOMA-SSK, and NOMA-GSSK. NOMA-SSK and NOMA-GSSK show better spectral efficiency than MIMO-NOMA because of the gain of exploiting the spatial domain. In addition, NOMA-GSSK has better spectral efficiency than NOMA-SSK, if the same number of transmit antennas is used, depending on the characteristics of GSSK. This clearly shows the comparison between NOMA-SSK, with $M_t=8$, and NOMA-GSSK, with $M_t=8, M_a=4$.
\par
Fig.~\ref{fig:EE} shows the energy efficiency comparison of MIMO-NOMA, NOMA-SSK, and NOMA-GSSK. In Fig. \ref{fig:EE}, we can see that the energy efficiency of NOMA-SSK with $M_t = 4$ and NOMA-GSSK with $M_t = 4, M_a = 2$ is the same, because the same energy is assigned, and the data rate is equal to 2 bps. However, when the number of transmit antennas is 8, the efficiency of NOMA-GSSK is better than that of NOMA-SSK, because the achievable data rate of NOMA-GSSK with $M_t=8, M_a=3$ is 4 bps. The achievable data rate of NOMA-SSK with $M_t=8$ is 3 bps. In this case, NOMA-GSSK can transmit one more bit at the one-channel bandwidth using the same amount of transmit energy.
\par
Fig.~\ref{fig:BER} shows the BER comparison of cell-edge users in MIMO-NOMA, NOMA-SSK, and NOMA-GSSK for the same number of bpcu case. The total number of users was 3. Therefore, three users were multiplexed in the power domain in MIMO-NOMA, and two users were multiplexed in the power domain in NOMA-SSK and NOMA-GSSK. Because NOMA-SSK and NOMA-GSSK utilize the spatial domain, interference caused by power allocation does not occur. As a result, their BER performance is better than that of MIMO-NOMA.

\section{Conclusion}
In this letter, we propose NOMA-GSSK using multiple active transmit antennas for performance enhancement. In NOMA-GSSK, the spatial domain was assigned to cell-edge users for transmitting symbol information using only the antenna index without SIC. In addition, computational complexity was reduced. Both analytical and simulation results show that the proposed scheme achieves significant spectral and energy efficiency gain, and BER and complexity reduction, compared to MIMO-NOMA.

\bibliographystyle{IEEEtran}
\bibliography{reference}

\end{document}